\documentclass[aps,prx,onecolumn,superscriptaddress,preprint]{revtex4-1}
\usepackage[english]{babel}
\PassOptionsToPackage{english}{babel}
\usepackage{lineno}

\usepackage[utf8]{inputenc}
\usepackage{color}
\usepackage{amsmath,amssymb,graphicx,xcolor,units,hyperref,subcaption}
\usepackage{booktabs}
\usepackage{soul}
\usepackage{pdfpages}
\usepackage{etoolbox}

\makeatletter
\patchcmd{\@outputpage@head}{\@ifx{\LS@rot\@undefined}{}{\LS@rot}}{}{}{}
\makeatother

\begin{document}
\selectlanguage{english}
\title{Heat rectification via a superconducting artificial atom}

\author{Jorden Senior}
\affiliation{QTF Centre of Excellence, Department of Applied Physics, Aalto
University School of Science, P.O. Box 13500, 00076 Aalto, Finland}

\author{Azat Gubaydullin}
\affiliation{QTF Centre of Excellence, Department of Applied Physics, Aalto
University School of Science, P.O. Box 13500, 00076 Aalto, Finland}

\author{Bayan Karimi}
\affiliation{QTF Centre of Excellence, Department of Applied Physics, Aalto
University School of Science, P.O. Box 13500, 00076 Aalto, Finland}

\author{Joonas T. Peltonen}
\affiliation{QTF Centre of Excellence, Department of Applied Physics, Aalto
University School of Science, P.O. Box 13500, 00076 Aalto, Finland}

\author{Joachim Ankerhold}
\affiliation{Institute for Complex Quantum Systems and IQST, University of Ulm, 89069 Ulm, Germany}

\author{Jukka P. Pekola}
\affiliation{QTF Centre of Excellence, Department of Applied Physics, Aalto
University School of Science, P.O. Box 13500, 00076 Aalto, Finland}

\maketitle
{\bf
In miniaturising electrical devices down to nanoscales, heat transfer has turned into a serious obstacle but also potential resource for future developments, both for conventional and quantum computing architectures\cite{Paolo,Paul,karimi_otto_2016}. Controlling heat transport in superconducting circuits has thus received increasing attention in engineering microwave environments for circuit quantum electrodynamics (cQED)\cite{wallraff,Matti,tan_quantum-circuit_2017} and circuit quantum thermodynamics experiments (cQTD)\cite{pekola2015,vinjanampathy_quantum_2016}. While theoretical proposals for cQTD devices are numerous\cite{martinezheat,Hwang,Drevillon,Pereira,Barzanjeh,Li,kosloff}, the experimental situation is much less advanced. There exist only relatively few experimental realisations\cite{Zettl,GiazottoRect, QHV,rossnagel_single_2016,Scheibner}, mostly due to the difficulties in developing the hybrid devices and in interfacing these often technologically contrasting components.
Here we show a realisation of a quantum heat rectifier, a thermal equivalent to the electronic diode, utilising a superconducting transmon qubit coupled to two strongly unequal resonators terminated by mesoscopic heat baths. Our work is the experimental realisation of the spin-boson rectifier proposed by Segal and Nitzan~\cite{Segal}.

}

A rectifier is a device in which the transport is directionally impeded, optimally to allow it only in one direction. In the charge regime, the ubiquitous rectifier, or diode, is one of the most fundamental components in electronic circuits, and can be realised relatively simplistically, for example as a device exploiting the depletion zone between electrons and holes in a semiconductor p-type/n-type junction.

The implementation of devices for manipulating charge current is enabled by the discrete and polarised nature of the carrier, and their interaction with electromagnetic fields. Heat currents do not have this quality, and manipulating their flow usually relies on precise control of the energy population distributions of the various materials involved in the device. In the superconducting regime, it is provided by the superconducting/semiconducting gap in one direction, and/or asymmetric couplings of the heat baths to quasiparticle heat carriers, for example by using metals with differing electron-phonon couplings.


The two-level system of a transmon-type qubit coupled to two unequal resonators is a well-placed tool for studying asymmetric photonic transport, each element having engineered resonances and couplings to each other that can be designed for various modes of operation~\cite{schmidt_photon-mediated_2004, meschke}. In particular, it represents a minimal set-up to explore, under well-controlled conditions, the subtle phenomenon of heat rectification which requires both non-linearities and symmetry breaking~\cite{Segal,Motz}. By utilising a superconducting quantum interferometer (SQUID) as the non-linear element of the transmon,
the Josephson energy $E_{J}(\Phi) \simeq E_{J0} \left| \cos(\pi \Phi/\Phi_0) \right|$ can be tuned by an incident magnetic flux $\Phi$. Here, $\Phi_0 = h / 2e$
is the magnetic flux quantum. This in turn allows to control the excitation frequencies of the transmon determined by $E_J(\Phi)$ and the charging energy $E_C$ with the transition frequencies $\omega_{n,n+1}$ between levels $n$ and $n+1$ ($n=0,1,2,...$) given by
\begin{equation}
	\omega_{n,n+1}(\Phi)=\omega_{\rm p}(\Phi)-(n+1)E_{C}/\hbar\, ,
	\label{eqn:qubit}
\end{equation}
where the plasma frequency $\omega_{\rm p}(\Phi)=\sqrt{8E_{C}E_{J}(\Phi)}/\hbar$. The $n$-dependence in Eq. \eqref{eqn:qubit} produces here the necessary non-linearity for rectification; this is the property that in fact makes the transmon a qubit in cQED applications. On the other hand, the symmetry breaking is due to the difference in effective qubit-bath couplings $g_{1}$ and $g_{2}$, via the left and right resonators of clearly different frequencies, $\omega_1<\omega_2$, respectively. This depends upon the spectral overlap between the corresponding resonator and the transmon and can be controlled by the applied flux~ \cite{koch_charge-insensitive_2007}.
Assuming one of the temperatures is $T=1/(k_{\rm B}\beta) > 0$ and the other one to be $=0$, or vice versa, the rectification ratio $\mathcal{R}$ of power $P_i$ to bath $i$ in the forwards ($+$) and backwards ($-$) direction \cite{Motz} is obtained in the two level approximation for the transmon as
\begin{equation}
\mathcal{R} =  \frac{ \lvert P^{+}_{i} \lvert }{ \lvert P^{-}_{i} \lvert} = \frac{g_{1} + g_{2}\coth(\frac{\beta \hbar \omega_{01}}{2})}{g_{1}\coth(\frac{\beta \hbar \omega_{01}}{2}) + g_{2}}\, .
\label{eqn:RectDef}
\end{equation}
Any value $\mathcal{R}\neq 1$ corresponds to heat rectification while $\mathcal{R}=1$ describes completely symmetric heat flux. By introducing the asymmetry in coupling factors $\delta = 1 - g_{1}/g_{2} $, this expression can be simplified for $\lvert \delta \rvert \ll 1 $ to read
\begin{equation}
	\mathcal{R} = 1 + e^{-\beta\hbar\omega_{01}}\delta\, .
	\label{eqn:RectShort}
\end{equation}

 In contrast, if one considers a scheme in which the superconducting qubit is replaced by a harmonic oscillator or with a single-level quantum dot, the linearity or fermionic nature of such a device instead does not lead to rectification (see supplementary material).


Figure \ref{fig:chip} presents a diagram of the realised device, consisting of a transmon qubit coupled to two resonators, designed at 2.8 GHz and 6.7 GHz frequencies, respectively. Each resonator is shunted at the current maximum with a copper thin-film resistor, shown in the left inset as a colourised scanning electron micrograph. This element provides the thermal bath, which is heated and measured by superconducting aluminium probes, grown using an electron beam evaporator with the Dolan bridge technique. The thin insulator is formed by the native oxide of aluminium, grown by an in-situ oxidation prior to the deposition of the copper. The SQUID of the transmon qubit is shown on the right inset, utilising a similar fabrication procedure. A simple diagrammatic model for understanding the system is also presented, with the role of the resonator-qubit-resonator structure represented by a diode.

Due to the definition of resistance being dissipative, normal metal baths used in this way as terminations to superconducting resonators lead to a substantial decrease in the quality factor of the resonator, from an intrinsic value in the range $10^4$, to a value of order $10$, as verified in the measurement in Ref.~\cite{Randy}. Lower resistance baths would result in higher quality resonators with larger asymmetry in the qubit-resonator coupling, and thus stronger rectification, but at the expense of lower power transfer.

Temperature differences between the normal metal baths can be controllably achieved by local Joule heating of each bath independently, with the tendency to thermally populate their corresponding resonator. We define the ``forward'' to be the direction from the low frequency (2.8 GHz) to high frequency (6.7 GHz) resonator, and declare the heated bath to be the source with the heat flowing to the target bath. Identical but opposite temperature bias can be applied in the ``reverse'' direction. The measured power on the target bath is shown in Figure \ref{fig:curves}a for various source temperatures between 380 mK and 420 mK. The horizontal axis depicts the static magnetic flux $\Phi$ on the transmon qubit, normalised with respect to the flux quantum $\Phi_0$, applied to tune its transition frequency.

We observe qualitatively different magnetic flux-dependences of the heat transport based upon the directionality. Domains of strong differences between forward and backward heat transfer with pronounced sub-structures alternate with those of weak discrepancies and relatively smooth traces. To better understand this directionality, one can extract the rectification by extracting the ratio of transmitted power in the forward direction to the reverse direction, see Figure \ref{fig:curves}b. The flux-tunable rectification is isolated by a subtraction of the non-tunable contribution $\mathcal{R}_{min}$. The origin for non-zero $\mathcal{R}_{min}$ 
is likely to originate from the uncertainty of the temperature bias, which we estimate to be $\pm$ 5 mK. This can lead to shifts in the non-flux dependent heat transport of up to 5 fW.

\begin{figure}
 \centering
 \includegraphics[width=\linewidth]{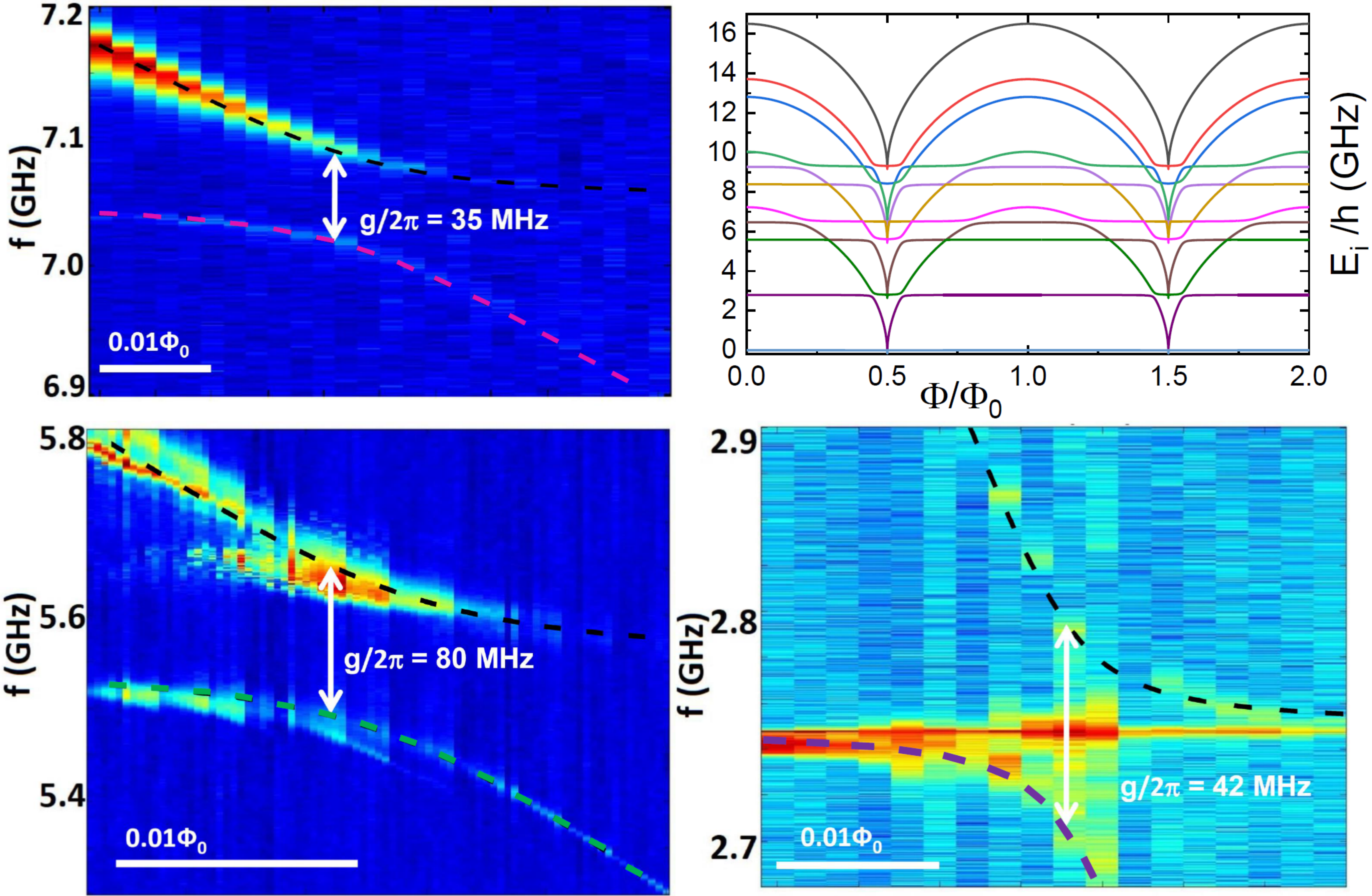}
  \caption{Two-tone spectroscopic readout of resonator-qubit-resonator structure performed using a tertiary readout resonator coupled to a third electrode of the qubit. Here, the copper terminations are not present. We observe qubit-resonator couplings at 2.75 GHz, corresponding to the low frequency resonator, 5.5 GHz, corresponding to the second mode of this resonator, and at 7.05 GHz, corresponding to the high frequency resonator. We note, however, that there is broadening of these lines when the copper termination is present, as reported in \cite{Randy}. The parameters used in the calculated energy spectra, shown in the upper right figure, are: $E_{\rm J}/h=45$ GHz and $E_{\rm C}/h=0.15$ GHz, which give $\omega_{01}(\Phi=0)/2\pi=7.2$ GHz.  }
  \label{fig:spectroscopy}
\end{figure}

Apparently, rectification appears to be almost independent of applied power (temperature gradient) but strongly depending on magnetic flux. The first behavior can be understood from the simple two state model discussed above. Equation (\ref{eqn:RectShort}) tells us that changes in the rectification coefficient between two different temperature gradients $T$ and $T+\Delta T$ with $\frac{\Delta T}{T}\ll 1$ are suppressed at low temperatures, namely, $\mathcal{R}(T)/\mathcal{R}(T+\Delta T)\approx 1- \frac{\Delta T}{T} \, \delta \beta\hbar\omega_{01}{\rm e}^{-\beta \hbar\omega_{01}}$. This argument also applies to more complex structures. The flux dependence is directly related to the energy levels of the resonator-qubit-resonator compound due to the prevailing resonator-qubit coupling compared to the resonator-bath interaction. Figure \ref{fig:spectroscopy} displays these levels of the measured structure using realistic parameters from independent measurements, plotted against $\Phi/\Phi_0$. We use the Hamiltonian and the basis states given in the supplementary material. Domains of more regular behavior in the spectrum alternate with those of high densities of avoided crossings when the flux is tuned. The latter appear at half-integer values of $\Phi/\Phi_0$. Consequently, we expect to observe a weak flux dependence of rectification away from these domains. The strong flux dependence near these degeneracies can be directly attributed to the fast varying transition spectrum of the transmon. Physically, heat transfer occurs via photon transfer between hot and cold reservoirs with the resonator-qubit-resonator structure acting as nonlinear medium. More specifically, since the resonators' frequencies are strongly de-tuned far beyond their respective linewidths, finite transmission is favoured if $\omega_{01}$ is tuned inbetween $\omega_1$ and $\omega_2$. In this situation, heat rectification is extremely sensitive to the transmon's level structure. While the first transition frequency Eq.~(\ref{eqn:qubit}) changes relatively smoothly with applied flux (see Figure \ref{fig:spectroscopy}), higher lying levels experience avoided crossings in narrow ranges of flux. In these regimes the spectral overlap with only one of the resonators changes abruptly, thus giving rise to pronounced peaks in the rectification. In this sense, the variations in the rectification coefficient can also be understood as the opening and closing of photonic transfer channels of a nonlinear medium between hot and cold reservoirs by varying the magnetic flux. Experimental spectra of the resonator-qubit-resonator structure by a two-tone measurement are shown in this figure as well. They demonstrate the higher frequency resonator at $\sim 7$~GHz, the second mode of the low frequency resonator at $\sim 5.5$~GHz, and similarly the lowest level of the low frequency resonator at $\sim 2.75$~GHz, all interacting with the qubit. The coupling energy of the resonators and the qubit is of order $50$~MHz given by the seperation at the anticrossing of the modes. The quality factor of the resonator, limited by the bath-phonon coupling, is estimated to be of order 10, based on measurements performed in \cite{Randy}.

In conclusion, we present wireless flux-tunable thermal rectification up to 10\% via the photon channel in a superconducting artificial atom. This phenomenon arises due to the anharmonicity of the system and its asymmetric coupling to two microwave
resonators, thermally populated by mesoscopic resistive baths. The device can be integrated with existing superconducting circuit architectures, in particular with superconducting qubits and Josephson logic, with potential applications in directionally manipulating heat flow in superconducting devices, for example in the fast initialisation of a qubit. Additionally, this device may be used as a platform for exploring coherent caloritronics, and the frontiers of quantum thermodynamics.


\subsection{Methods}
\emph{Fabrication protocols} \\
These devices are fabricated on a high resistivity silicon wafer, upon which a $20\,\mathrm{nm}$-thick alumimina film has been grown by atomic layer deposition, followed by a susbsequently sputtered $200\,\mathrm{nm}$-thick niobium film.
Larger features, namely the microwave resonators (measured spectroscopically at $2.8\,\mathrm{GHz}$ and $6.7\,\mathrm{GHz}$, see Figure \ref{fig:chip}), transmon island, and superconducting probe fanout are patterned by electron beam lithography, that is transferred to the niobium by reactive ion etching utilising fluorine chemistry.
The coplanar waveguide resonator has characteristic impedance $50\,\mathrm{\Omega}$, formed of a $20\,\mathrm{\mu m}$-wide centreline spaced by $10\, \mathrm{\mu m}$ with respect to the ground plane.
The tunnel junction elements in the superconducting probes and interferometer are also patterned by electron beam lithography, then transferred onto the wafer by physical vapour deposition of aluminium in an electron beam evaporator with an \textit{in situ} oxidation step, resulting in a per junction resistance of $22 \pm 3 \,\mathrm{k\Omega}$ and $6 \pm 3 \,\mathrm{k\Omega}$ for the Normal-Insulator-Superconductor probes and superconducting interferometer, respectively.
To ensure sufficient contact between lithographic layers, specifically between the niobium and aluminium features, an \textit{in situ} Ar ion plasma milling is performed before the aluminium deposition process step.
After liftoff in acetone and cleaning in isopropyl alcohol, the substrates are diced to $7 \times 7 \, \mathrm{mm}$ size by diamond-embedded resin blade and wire-bonded to a custom made brass stage.

\begin{figure}[th]
	\centering
	\includegraphics [width=0.8\columnwidth] {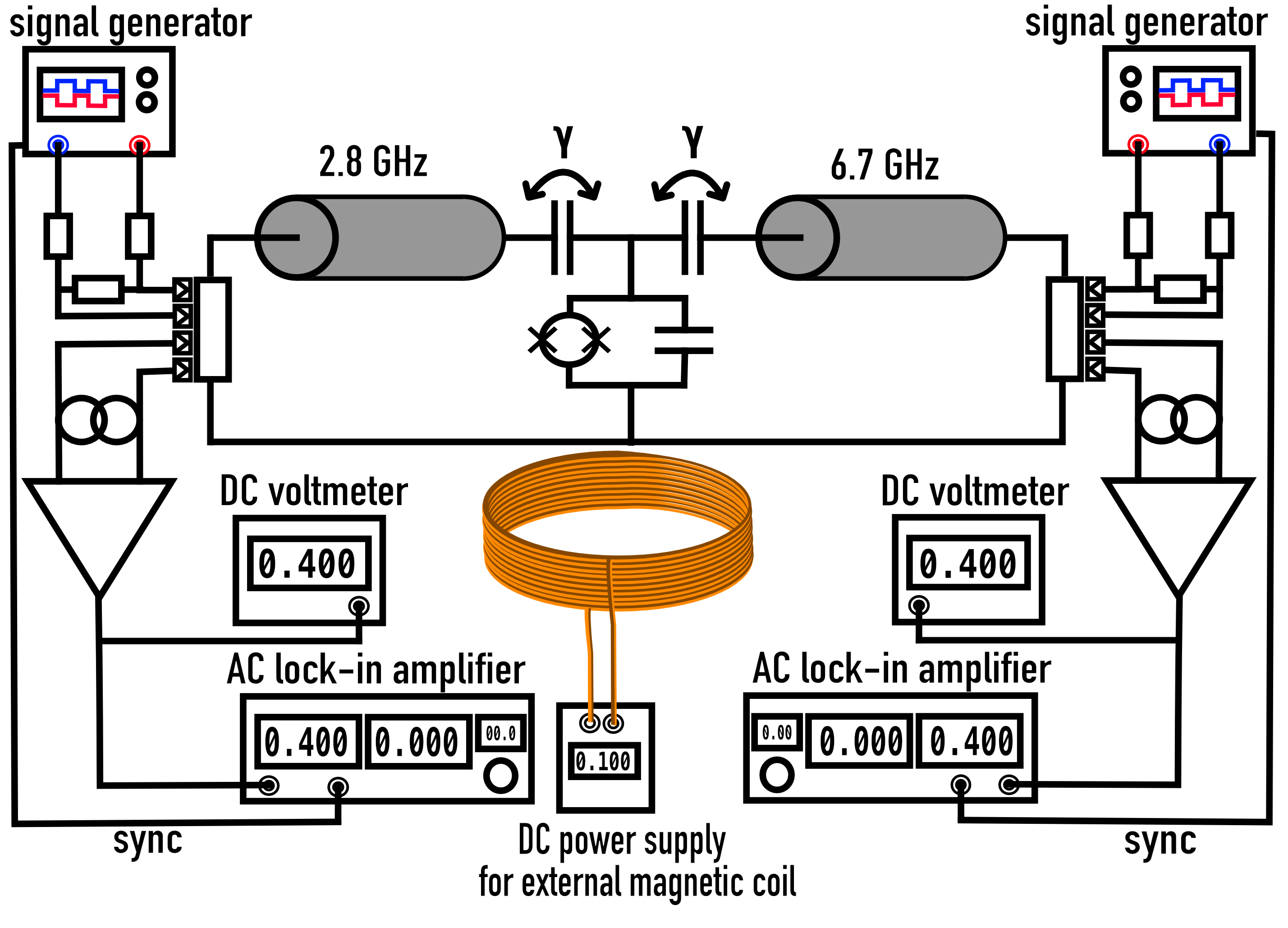}
	\caption{Schematic of the measurement set-up.
		\label{fig_exp}}
\end{figure}

\emph{Measurements} \\
The measurements are performed in a custom-made plastic dilution refrigerator with the cooling stage set at $150\, \mathrm{mK}$ for these measurements.
The bonded chip is enclosed by two brass Faraday shields, and is
readout at room-temperature via $1\,\mathrm{m}$ of Thermocoax filtered cryogenic lines, resulting in an effective signal bandwidth of $0-10\, \mathrm{kHz}$, for low-impedance loads.
The magnetic-flux-tuning of the energy level spacing of the transmon is achieved by a superconducting solenoid encompassing the entire sample stage assembly, inside of a high-permeability magnetic shield. It is mounted inside the refrigerator at a temperature of $4 \, \mathrm{K}$.
Electronic characterisation is applied by programmable function generators at room temperature and measured by a low noise amplification chain (room-temperature low-noise
amplifiers FEMTO Messtechnik GmbH DLPVA-100) into a DC multimeter (for DC measurements), and additionally into a lock-in amplifier synchronised to the square-wave modulation ($22\, \mathrm{Hz}$)
of the voltage bias of the heated thermal reservoir, in the configuration shown in Figure \ref{fig_exp}.
By applying a current bias between the superconducting probes close to the superconducting energy gap, hot electrons from the normal metal can tunnel into the superconducting probes, and a voltage signal can be measured, with a well characterised dependence on the temperature of the electrons in the metal in the range 100 mK - 500 mK, of interest to this experiment. By assuming that the heat is well localised due to the near-perfect Andreev mirrors at the superconductor-normal metal interface, and that the main relaxation channel for the heat is via the electron-phonon coupling, the power can be estimated by $P_\mathrm{el-ph} = \Sigma \mathcal{V} (T_\mathrm{el}^5 - T_\mathrm{ph}^5)$, where $\Sigma$ is the material dependent electron-phonon coupling constant, $\mathcal{V}$ is volume of the normal metal bath, and $T_\mathrm{el}$ and $T_\mathrm{ph}$correspond to the electron temperature and phonon temperature respectively. As the experiment is performed under steady-state conditions, we can assume that the phonon temperature is in equilibrium with the cryostat base temperature, measured by a ruthenium oxide thermometer that has been calibrated against a Coulomb blockade thermometer. Additionally, by voltage biasing the SINIS structure sufficiently above the superconducting energy gap, Joule heating of the baths can be applied (for small voltage biasing below the gap, evaporative cooling can also occur). Hence, by utilising four superconducting probes, we can both engineer the temperature of the baths, and measure this induced temperature in parallel \cite{revmodgiaz}.

\section{Acknowledgements}
This work was funded through Academy of Finland grant  312057
and from the European Union's Horizon 2020
research and innovation programme under the European Research Council (ERC)
programme and Marie Sklodowska-Curie actions (grant agreements 742559 and
766025). J.A. acknowledges financial support from the IQST and the German Science Foundation (DFG) under AN336/12-1 (For2724).
This work was supported by Quantum Technology Finland (QTF) at Aalto
University.
We acknowledge the facilities and technical support of Otaniemi research
infrastructure for Micro and Nanotechnologies  (OtaNano), and VTT technical
research centre for sputtered Nb films.

We acknowledge L.B. Wang for technical help and thank Y.-C. Chang, A. Ronzani, D. Golubev, G. Thomas, and R. Fazio for useful
discussions.

\section{Author contributions}
J.S. and A.G. designed, fabricated, and measured the samples. Modelling of the work which is detailed in the supplementary material was done by B.K. and J.P.P. Technical support in fabrication, low-temperature setups, and measurements were provided by J.T.P. All authors have been involved in the analysis, and discussion of scientific results and implications of this work. The manuscript was written by J.S., B.K., J.P.P. and J.A.

\section{Competing financial interests}
The authors declare no competing financial interests.

\section{Data availability}
The data that support the plots within this article
are available from the corresponding author upon reasonable request.

\newpage

\begin{figure}
	\includegraphics[
  width=10cm,
  keepaspectratio]{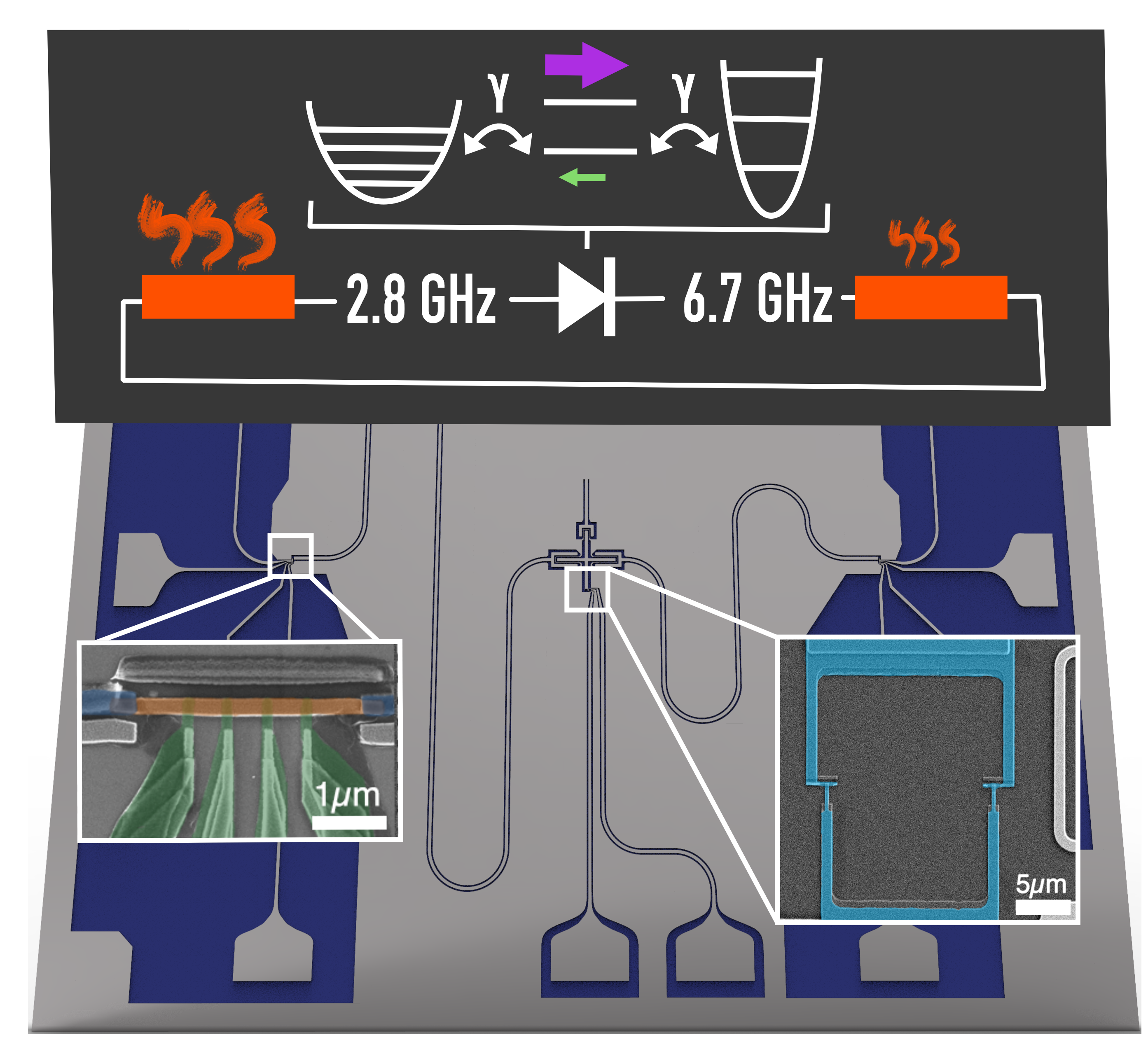}
	\caption{Diagram of the sample, consisting of a centrally located Xmon type superconducting qubit cross-coupled to two superconducting co-planar waveguide resonators at 2.8 GHz and 6.7 GHz, respectively. Each resonator is terminated with a thin-film copper microstrip resistor, acting as a mesoscopic thermal bath, one of which shown in the left inset by a colourised scanning electron micrograph of the copper microstrip resistor (orange),  with 4 superconducting aluminium probes (green) (separated from the copper by an insulator, not visible) for temperature control and readout, and two superconducting aluminium contacts to the co-planar waveguide resonator (blue). The superconducting quantum interferometer is shown similarly on the right inset. The topmost diagram represents a simple model of the system, with the resonator-qubit-resonator structure represented as a diode, and the forward and reverse directions drawn in purple and green respectively. A third electrode can be seen on the Xmon island, which in non-dissipative variants of the device used for spectroscopy, connects the Xmon to a readout resonator}
	\label{fig:chip}
\end{figure}

\begin{figure}[ht]
	\includegraphics[width=\linewidth]{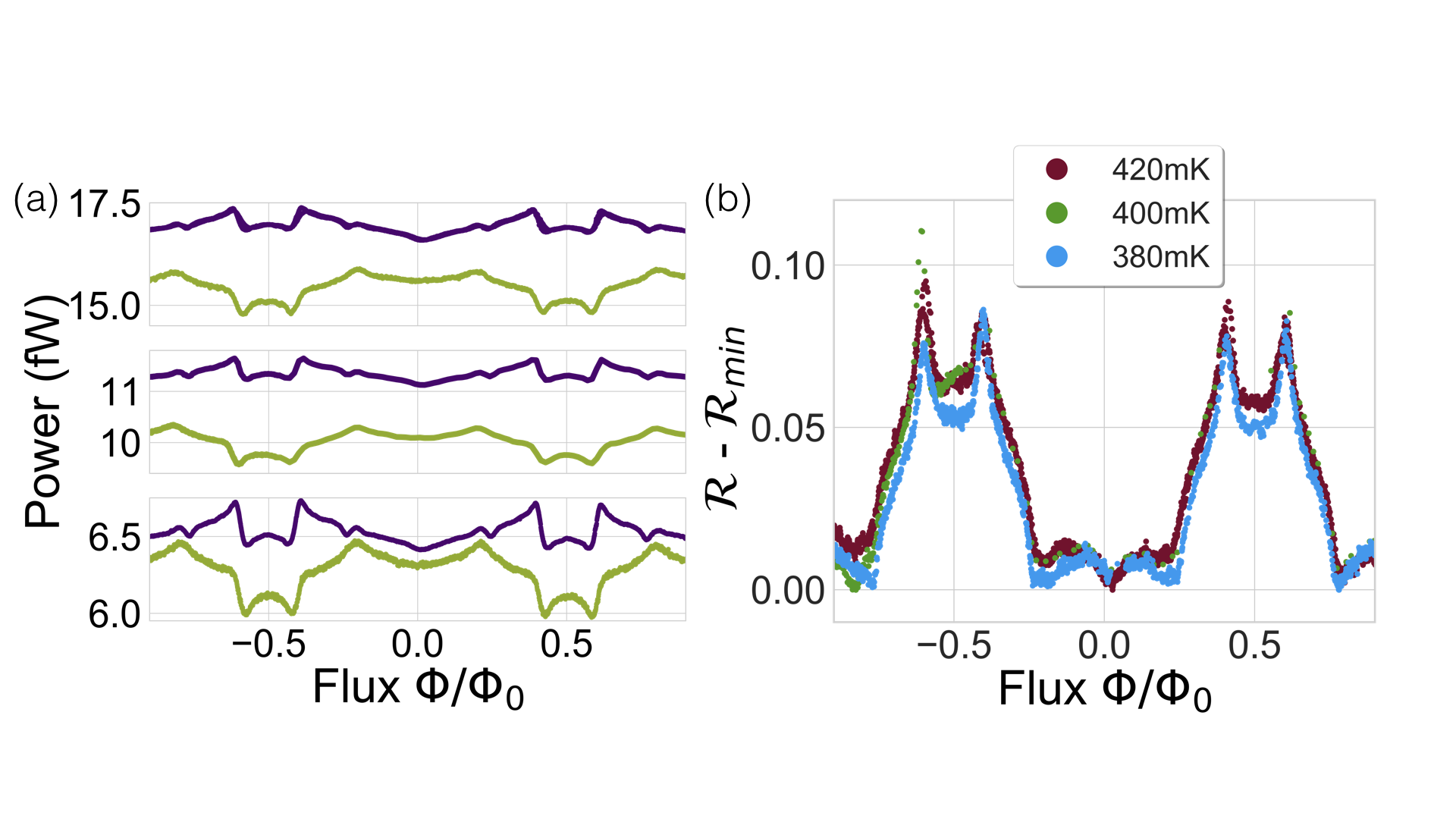}
	\caption{(a)Power transmitted between the two baths at three voltage (heating) bias points, with the subplots corresponding to 420 mK (1000 fW), 400 mK (750 fW), and 380 mK (600 fW) source temperatures (powers), in descending order. The bath temperature is kept fixed at 150 mK. Purple is the forward direction, and green the reverse, as shown in Figure \ref{fig:chip}. (b) Rectification ratio of traces from (a), with the non-tunable contribution removed.}
	\label{fig:curves}
\end{figure}




\newpage

\newpage
\includepdf[pages=1]{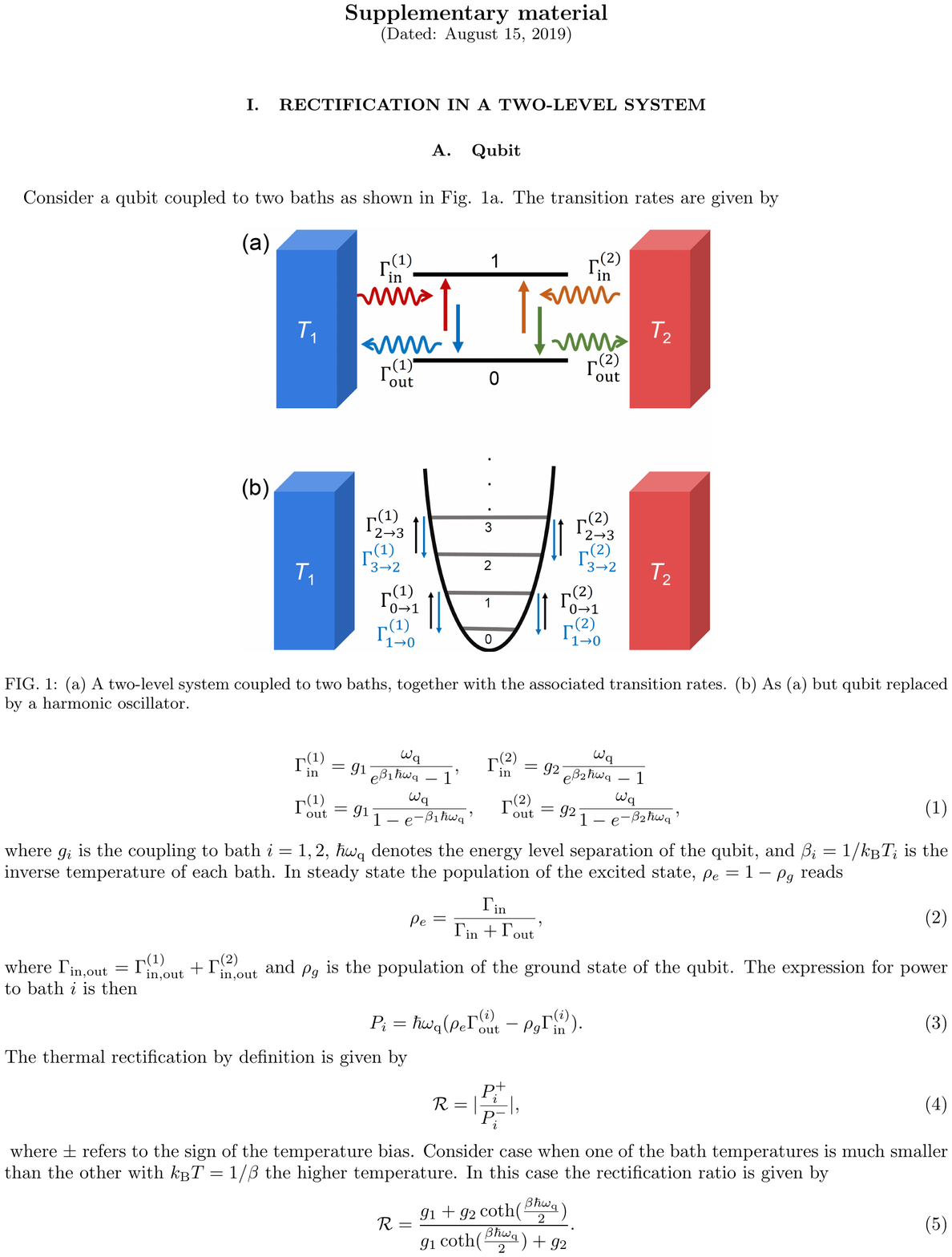}
\includepdf[pages=2]{HeatRect_SI.pdf}
\includepdf[pages=3]{HeatRect_SI.pdf}
\includepdf[pages=4]{HeatRect_SI.pdf}
\end{document}